\documentstyle[12pt]{article}
\begin{document}
\input epsf
\newcommand{\A}{{\mathcal{A}}}
\newcommand{\dA}{\delta{\mathcal{A}}}
\newcommand{\Od}{{\cal O}}
\newcommand{\lsim}   {\mathrel{\mathop{\kern 0pt \rlap
  {\raise.2ex\hbox{$<$}}}
  \lower.9ex\hbox{\kern-.190em $\sim$}}}
\newcommand{\gsim}   {\mathrel{\mathop{\kern 0pt \rlap
  {\raise.2ex\hbox{$>$}}}
  \lower.9ex\hbox{\kern-.190em $\sim$}}}
\begin{center}
\large{Essay written for the Gravity Research Foundation 2009
Awards for Essays on Gravitation} \\
\vspace{2cm}
\Large{\bf Dark energy: the absolute electric potential
of the universe} \\
\vspace*{1cm}
\large{\bf  Jose Beltr\'an Jim\'enez$^*$ and Antonio L. Maroto$^\dagger$}  \\
\vspace{0.3cm}
\normalsize
Departamento de F\'{\i}sica Te\'orica\\
Universidad Complutense de Madrid\\
28040 Madrid, Spain\\
\vspace{0.2cm}
March 30th, 2009

\vspace*{0.5cm} {\bf ABSTRACT}\\  \end{center} Is there an
absolute cosmic electric potential? The recent discovery of the
accelerated expansion of the universe could be indicating that
this is certainly the case. In this essay we show that the
consistency of the covariant and gauge invariant
 theory of electromagnetism is truly questionable when considered
on cosmological scales.
Out of the four
components of the electromagnetic field,
Maxwell's theory  only contains two physical
degrees of freedom.  However, in the presence of gravity, one of the
 "unphysical" states cannot be consistently eliminated,
thus  becoming real. This third polarization state is completely
decoupled from charged matter, but can be excited gravitationally thus
breaking gauge invariance.
On large scales the new state can be seen as a
homogeneous cosmic electric potential, whose energy density behaves as a
cosmological constant.

\vspace*{0.5cm}
\noindent
\rule[.1in]{4.5cm}{.002in}

\noindent $^*$ jobeltra@fis.ucm.es \\
\noindent $^\dagger$ maroto@fis.ucm.es

\newpage
\baselineskip 20pt

The recent discovery of the accelerated expansion of the universe, and
the difficulties found in the context of General Relativity (GR) and
the Standard
Model (SM) of elementary particles to properly account for this effect,
has led
to consider the possibility that physics on large scales could differ
from  our well-known small scale laws.

In this context, models in which the description of  the own
gravitational interaction are modified on large scales with
respect to GR have been extensively considered in recent years.
Here we will concentrate in the other long-range interaction of
nature, and explore the possibility that our standard theory of
(quantum) electromagnetism,  being valid on small scales, could
give rise to unexpected effects on cosmological scales. As a
matter of fact, this possibility is perfectly compatible with
current experimental limits which have tested electromagnetism
only for wavelengths roughly below the Solar System scales (1.3
A.U. \cite{Nieto}).

In this essay we will discuss one of the most striking
consequences of electromagnetism in the cosmological context, which
is the possibility that the universe at large scales not only sets
a privileged reference frame, but  could also determine an
absolute electric potential. Indeed, it is well known that the
presence of matter and radiation in the universe implies that, on
large scales, the universe as a whole has associated a privileged
reference frame. That frame is nothing but the {\it cosmic center
of mass frame} \cite{moving} of the different components (baryonic
and dark matter, radiation and dark energy). In the case in which
all such components are at rest with respect to each other, the
frame can be identified with that of the observers who see an
isotropic cosmic microwave background. Thus we can say that,
although   Lorentz symmetry is locally
 a good symmetry of space-time, it is broken on large scales
by the matter/energy content of the universe. But, what about the
rest of gauge symmetries and, in particular, that of
electromagnetism? Is it also possible that, although on small
scales we see electromagnetic gauge symmetry as an exact symmetry
of nature, the actual situation is that it is broken by the
content of the universe on large scales? Does it  make sense to
talk about a privileged electromagnetic gauge? We will argue that
dark energy, responsible for the accelerated expansion of the
universe, could be nothing but the energy density associated to
such absolute electric potential \cite{sector,EM}.

Let us start by briefly reviewing the standard covariant electromagnetic
quantization in Minkowski space-time \cite{Itzykson}.
 The starting point is the action:
\begin{eqnarray}
S=\int d^4x \left(-\frac{1}{4}F_{\mu\nu}F^{\mu\nu}+\frac{\xi}{2}
(\partial_\mu A^\mu)^2+
A_\mu J^\mu\right)
 \label{actionGB}
\end{eqnarray}
which is not invariant under general gauge transformations, but
only under residual ones, given by:
$A_\mu\rightarrow A_\mu+\partial_\mu \theta$, with
$\Box \theta=0$.
The equations of motion obtained from this action  read:
\begin{eqnarray}
\partial_\nu F^{\mu\nu}&+&\xi\partial^\mu(\partial_\nu A^\nu)=J^\mu
\label{fieldeq}
\end{eqnarray}
In order to recover ordinary Maxwell's equation, the Lorenz condition
$\partial_\mu A^\mu=0$ must be
imposed so that the $\xi$ term disappears. At the classical level
this can be achieved by means of appropriate boundary conditions
on the field. Indeed, taking the four-divergence of the above equation,
we find:
\begin{eqnarray}
\Box(\partial_\nu A^\nu)=0
\end{eqnarray}
where we have made use of current conservation. This means that
the field  $\partial_\nu A^\nu$ evolves as a free scalar field, so that
if it vanishes for large $\vert t \vert$, it will vanish for all time.
At the quantum level, the Lorenz condition cannot be imposed as an
operator identity, but only in the weak sense
$\partial_\nu A^{\nu \,(+)}\vert \phi\rangle=0$, where
$(+)$ denotes the positive frequency part of the operator and
$\vert \phi\rangle$ is a physical state.
This condition is equivalent to imposing that the physical  states
contain the same number of
temporal and longitudinal photons, so that their energy densities,
having opposite signs, cancel each other. Thus we see that the
Lorenz condition seems to be
essential in order to recover standard
Maxwell's equations and get rid of the negative energy states.

Now we move to an expanding universe.  The curved space-time version of
action (\ref{actionGB}) reads:
\begin{eqnarray}
S=\int d^4x \sqrt{g}\left(-\frac{1}{4}F_{\mu\nu}F^{\mu\nu}+\frac{\xi}{2}
(\nabla_\mu A^\mu)^2+
A_\mu J^\mu\right)
 \label{actionF}
\end{eqnarray}
and the modified Maxwell's equations are:
\begin{eqnarray}
\nabla_\nu F^{\mu\nu}+\xi\nabla^\mu(\nabla_\nu A^\nu)=J^\mu
\label{EMeqexp}
\end{eqnarray}
Taking again the four divergence, we get:
\begin{eqnarray}
\Box(\nabla_\nu A^\nu)=0\label{minimal}
\end{eqnarray}
We see that once again  $\nabla_\nu A^\nu$  behaves as a
scalar field which is decoupled from the conserved electromagnetic currents,
but it is non-conformally coupled to gravity. This means that, unlike
the flat space-time case,  this
field can be excited from quantum vacuum fluctuations by the expanding
background in a completely analogous way to the inflaton fluctuations
during inflation. Thus this poses the question of the validity
of the Lorenz condition at all times.

In order to illustrate this effect, we will present a toy example.
Let us consider quantization in the absence of currents,
in a spatially flat expanding background,
whose metric is written in conformal time as:
\begin{eqnarray}
ds^2=a(\eta)^2(d\eta^2-d\vec x^2)
\end{eqnarray}
For the scale factor we assume the following form:
$a(\eta)=2+\tanh(\eta/\eta_0)$ where $\eta_0$ is constant. 
This metric is asympotically 
flat in the remote past and far future. Let us prepare our 
system  in an initial  state $\vert \phi\rangle$ belonging 
to the physical Hilbert space, 
i.e. satisfying $\partial_\nu \A^{\nu \,(+)}_{in}\vert \phi\rangle=0$
in the initial flat region. 
We  solve the coupled system of equations
(\ref{EMeqexp}) for the corresponding Fourier modes $\A_{\mu\vec k}$. 
Because of the expansion of the universe, the positive frequency
modes in the $in$ region with a given temporal or longitudinal
polarization $\lambda$ will become a linear superposition of
positive and negative frequency modes in the $out$ region and with
different polarizations $\lambda'$ \cite{Birrell}.
\begin{figure}[h]
\begin{center}
{\epsfxsize=9.5cm\epsfbox{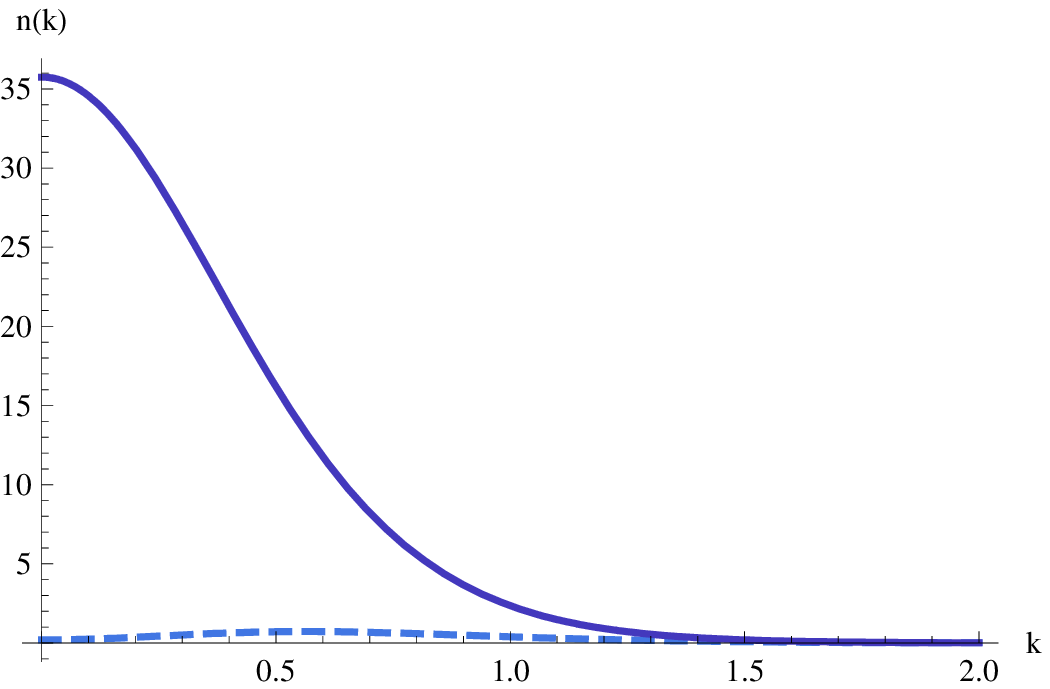}} \end{center}
{\footnotesize {\bf Figure 1:} Occupation numbers for temporal (continuous
line) and
longitudinal (dashed line) photons in the $out$ region vs.
$k$ in $\eta_0^{-1}$ units.}
\end{figure}

Thus, the system will end up in a final state which no longer
satisfies the weak Lorenz condition, i.e. in the out region
$\partial_\nu \A^{\nu \,(+)}_{out}\vert \phi\rangle\neq 0$.
This is shown in Fig. 1, where we have computed  the final number
of temporal and longitudinal photons starting from an
initial vacuum state with  $n_0^{in}(k)=n_\parallel^{in}(k)=0$.
We see that
in the final region $n_0^{out}(k)\neq n_\parallel^{out}(k)$. Notice that
the failure comes essentially from large scales ($k\eta_0\ll 1$), since
on small scales ($k\eta_0\gg 1$), the Lorenz condition can be restored.
Motivated by this fact in the following we will explore
the possibility of quantizing electromagnetism in an expanding
universe without imposing this condition.

Let us then assume that the fundamental theory for electromagnetism
is given by (\ref{actionF}) and is not invariant under general gauge 
transformations, but 
only under residual ones.
The general solution for the modified equations (\ref{EMeqexp})
can be written as:
\begin{eqnarray}
\A_\mu=\A_\mu^{(1)}+\A_\mu^{ (2)}+\A_\mu^{(s)}+\partial_\mu \theta
\end{eqnarray}
where $\A_\mu^{(i)}$ with $i=1,2$ are the two transverse modes of the
massless photon, $\A_\mu^{(s)}$ is the new scalar state, which is the mode that
would have been
eliminated if we had imposed the Lorenz condition and, finally,
$\partial_\mu \theta$ is a pure gauge mode which can be
eliminated. In order to quantize the free theory,
we perform the mode expansion for the {\it three} physical states:
\begin{eqnarray}
\A_{\mu}=\int d^3\vec{k}
\left[\sum_{\lambda=1,
2,s}\left({\bf a}_\lambda(k)\A_{\mu k}^{(\lambda)}
+{\bf a}_\lambda^\dagger(k)\overline{\A_{\mu k}^{(\lambda)}}
\right)\right]
\end{eqnarray}
In fact, the three modes can be chosen to have positive
normalization and,  for $\xi=1/3$, the canonical commutation
relations are satisfied:
\begin{eqnarray}
\left[{\bf a}_\lambda(\vec{k}),{\bf a}_{\lambda'}^\dagger(\vec{k'})\right]
=\delta_{\lambda\lambda'}\delta^{(3)}(\vec{k}-\vec{k'}),\;\;\;
\lambda,\lambda'=1,2,s
\end{eqnarray}
with positive sign for the
three physical states, i.e. there are no negative norm states
in the theory, which in turn guarantees that there are no
negative energy states (ghosts).
Moreover, as shown in \cite{EM,VT}, the theory does not exhibit
either local gravity inconsistencies or classical instabilities.

As shown in  (\ref{minimal}),  $\nabla_\mu\A^{\mu}$ evolves 
as a minimally coupled scalar field. This means that 
 on sub-Hubble scales ($\vert k\eta\vert \gg 1$), the field    
is suppressed by the universe expansion as 
$\vert \nabla_\mu\A^{(s)\mu}_k\vert \propto a^{-1}$.
Thus, on small scales, the modified 
Maxwell's equations (\ref{EMeqexp}) will be 
physically indistinguishable from the flat space ones.
To summarize, from the previous discussion we see that 
the theory is consistent even though we have not imposed the 
Lorenz condition.
But, moreover, 
 on super-Hubble scales ($\vert k\eta\vert \ll 1$), we find  
$\vert\nabla_\mu\A^{(s)\mu}_k\vert= const.$ which, as shown in \cite{EM},  
implies that
the field contributes as an {\it effective cosmological constant} in 
(\ref{actionF}).

In order to determine its value, we will assume that the field is generated during inflation
from quantum vacuum fluctuations, in a completely analogous way 
to cosmological metric perturbations. Thus, in an inflationary de Sitter 
space-time, it is possible 
to obtain the corresponding dispersion:
\begin{eqnarray}
\langle 0\vert(\nabla_\mu\A^{\mu})^2\vert 0 \rangle=\int\frac{dk}{k}P_A(k)
\end{eqnarray}
where 
$P_A(k)=4\pi k^3\vert\nabla_\mu\A^{(s)\mu}_k\vert^2 $.
In the super-Hubble limit, we obtain for the power-spectrum:
\begin{eqnarray}
P_A(k)=\frac{9H_I^4}{16\pi^2},
\end{eqnarray}
with $H_I$ the constant Hubble parameter during inflation. 
Thus the electromagnetic energy
density on cosmological scales is given by
$\langle 0\vert\rho_A\vert 0 \rangle\sim (H_I)^4$.
The measured value of the
dark energy density then requires $H_I\sim 10^{-3}$ eV, which corresponds
to an inflationary scale of $M_I\sim 1$ TeV.
Thus we see that the  cosmological constant value
can be naturally explained in terms of physics at the electroweak scale.

\begin{figure}[h]
\begin{center}
{\epsfxsize=9.5cm\epsfbox{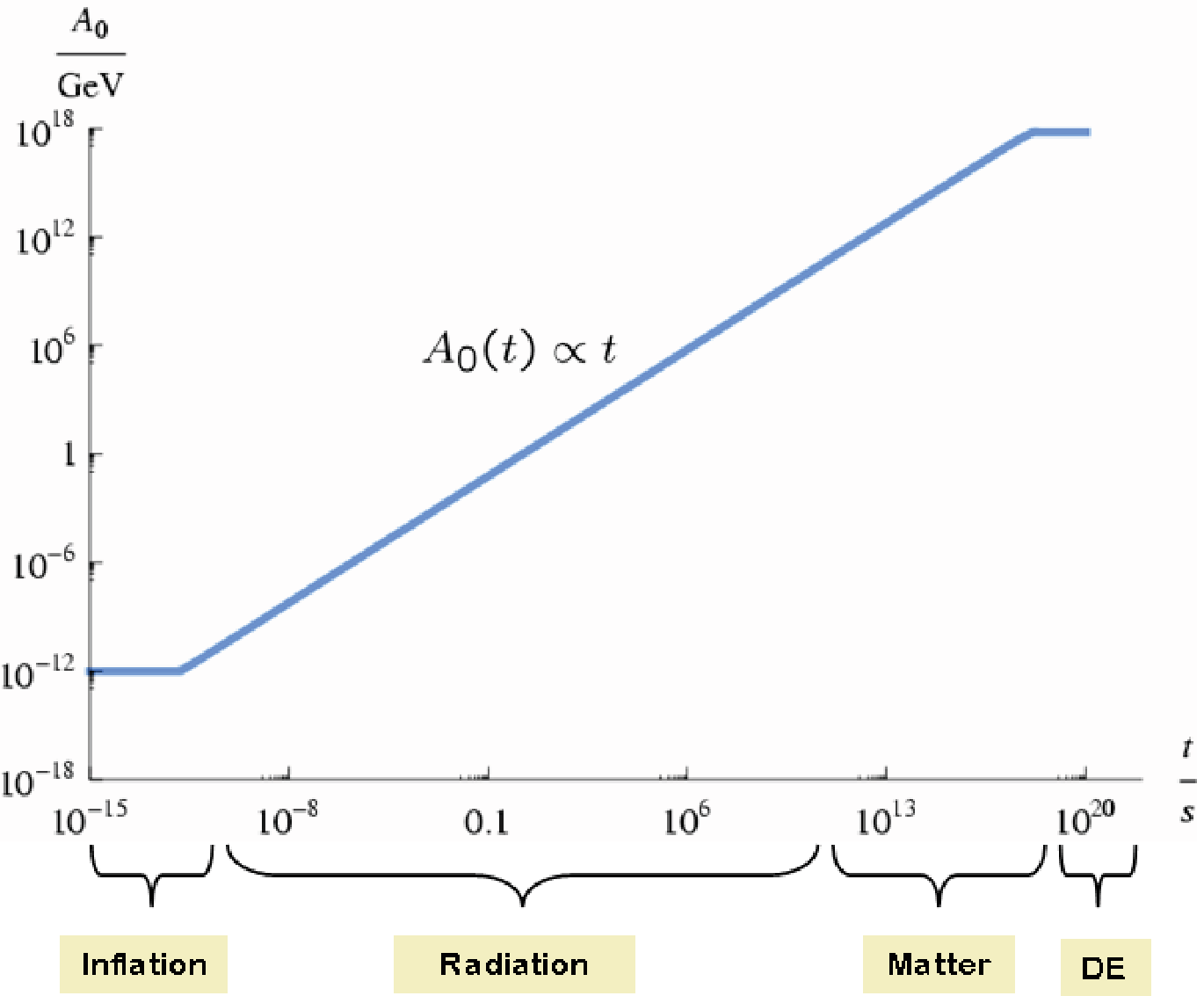}} \end{center}
{\footnotesize {\bf Figure 2:} Cosmological evolution of the electric
potential $A_0$ from electroweak-scale inflation until present.}
\end{figure}

Once the field is produced during inflation, its cosmological
evolution can be easily calculated \cite{EM}. We find that in
cosmological time
($dt=a(\eta)d\eta$), the homogeneous temporal component $A_0(t)$
(the electric potential)
is constant during inflation and grows as $t$ during matter and
radiation eras. When the electromagnetic dark energy   
 starts dominating, $A_0(t)$ becomes also constant. The
spatial components on super-Hubble scales $\vec A(t)$ are shown to grow
more slowly than $A_0(t)$ and can be neglected. In Fig. 2
we show the cosmological evolution of the electric potential
from its initial value generated in inflation ($A_0(t_I)\sim 10^{-3}$ eV)
up to its present value $A_0(t_0)\sim 0.3\, M_P$ with $M_P\sim 10^{19}$ GeV
the Planck mass.  

In conclusion, we have discussed the possibility
that the true theory of electromagnetism contains three and not
only two physical degrees of freedom.
 Although the new scalar
state is completely decoupled from the conserved currents, it 
can be gravitationally amplified during inflation, giving rise to
the observed dark energy density. 
The accelerated expansion of the
universe would be then the natural
consequence of the existence of an absolute electric potential
in the universe.

\vspace{0.2cm}

 {\em Acknowledgments:}
 This work has been  supported by
Ministerio de Ciencia e Innovaci\'on (Spain) project numbers
FIS 2008-01323 and FPA
2008-00592, UCM-Santander PR34/07-15875, CAM/UCM 910309 and
MEC grant BES-2006-12059.

\end{document}